\documentclass[prl,twocolumn,superscriptaddress,floatfix]{revtex4}
\usepackage{amsmath}
\usepackage{graphicx}
\usepackage{amssymb}
\usepackage{amsthm,amscd}
\usepackage{amsfonts}
\usepackage{subfigure}
\usepackage{psfrag}
\usepackage{times}
\usepackage{pstricks}
\usepackage{wasysym}

\newcommand{\dslash}{\not{\hbox{\kern-2pt $\partial$}}}
\newcommand{\bea}{\begin{eqnarray}}
\newcommand{\eea}{\end{eqnarray}}

\begin{document}

\title{Universality of liquid-gas Mott transitions at finite temperatures}
\author{Stefanos Papanikolaou}
\affiliation{Department of Physics, University of Illinois at Urbana-Champaign, 1110 W.
Green St., Urbana, IL 61801-3080}
\author{Rafael Monteiro Fernandes}
\affiliation{Ames Laboratory and Department of Physics and Astronomy, Iowa State
University Ames, IA 50011, USA}
\affiliation{Instituto de F\'{\i}sica "Gleb Wataghin", Universidade Estadual de Campinas,
and Laborat\'{o}rio Nacional de Luz S{\'{\i}}ncrotron,Campinas, SP, Brazil}
\author{Eduardo Fradkin}
\affiliation{Department of Physics, University of Illinois at Urbana-Champaign, 1110 W.
Green St., Urbana, IL 61801-3080}
\author{Philip W. Phillips}
\affiliation{Department of Physics, University of Illinois at Urbana-Champaign, 1110 W.
Green St., Urbana, IL 61801-3080}
\author{Joerg Schmalian}
\affiliation{Ames Laboratory and Department of Physics and Astronomy, Iowa State
University Ames, IA 50011, USA}
\author{Rastko Sknepnek}
\affiliation{Ames Laboratory and Department of Physics and Astronomy, Iowa State
University Ames, IA 50011, USA}
\date{\today }

\begin{abstract}
We explain in a consistent manner the set of seemingly conflicting
experiments on the finite temperature Mott critical point, and demonstrate
that the Mott transition is in the Ising universality class. We show that,
even though the thermodynamic behavior of the system near such critical
point is described by an Ising order parameter, the global conductivity can
depend on other singular observables and, in particular, on the energy
density. Finally, we show that in the presence of weak disorder the
dimensionality of the system has crucial effects on the size of the critical
region that is probed experimentally.
\end{abstract}

\maketitle

\begin{figure}[tb]
\centering
\subfigure[]{\includegraphics[width=0.15\textwidth]{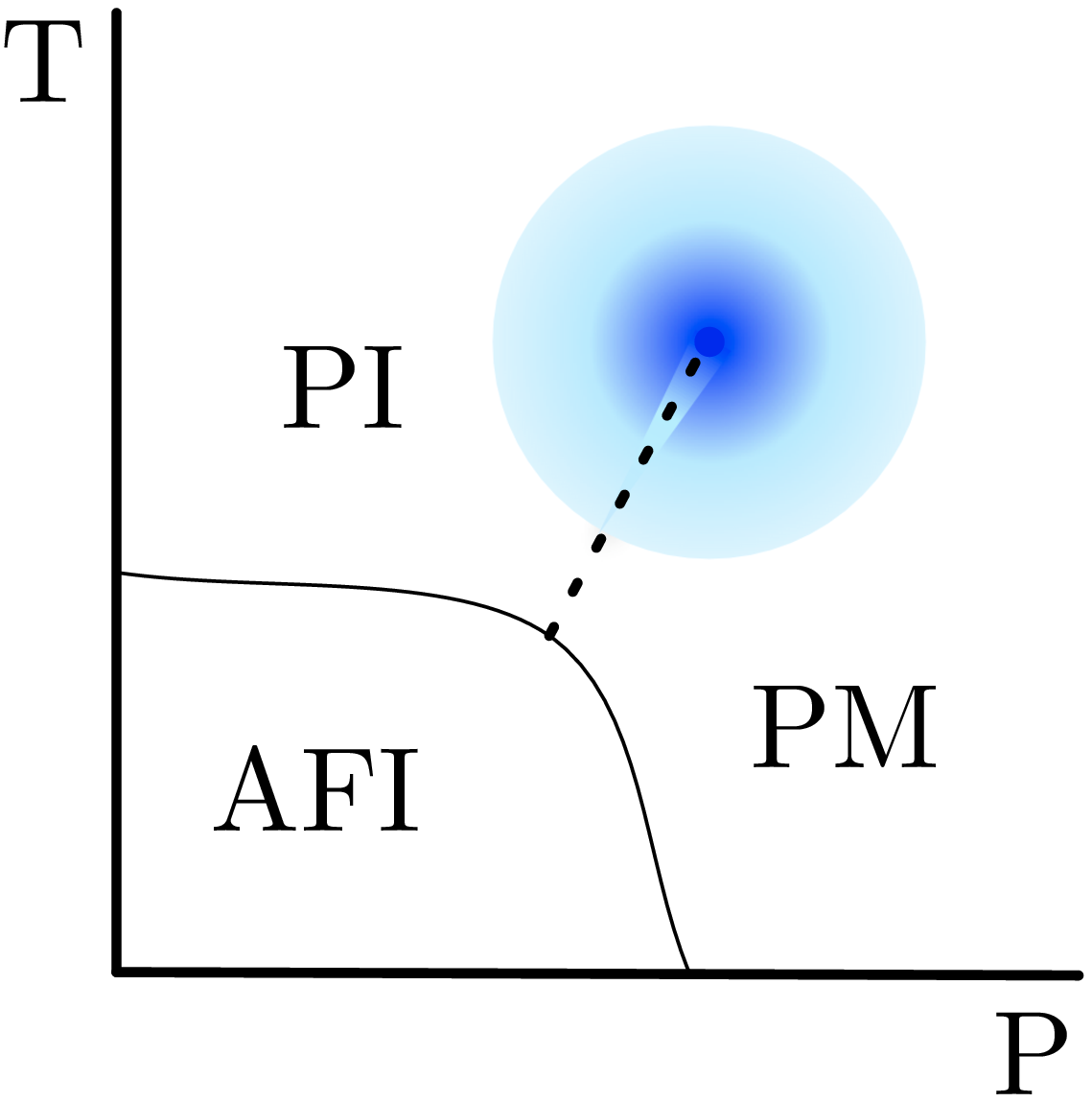}
\label{sketchypd}} \;\;\;\; \subfigure[]{\includegraphics[width=0.2
\textwidth]{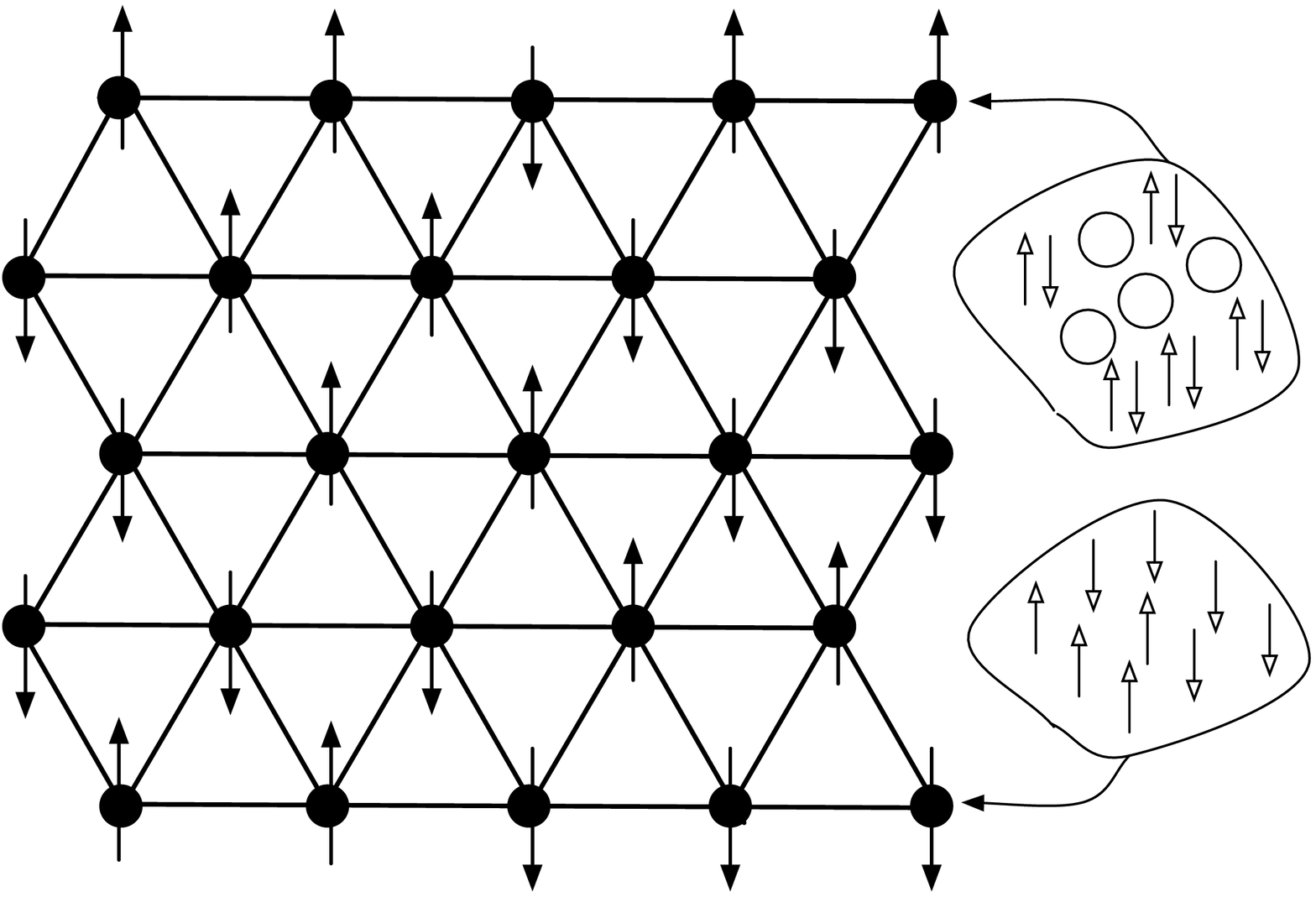}\label{lattice}}
\caption{(a) Typical phase diagram of Mott transitions as a function of
pressure $P$ and temperature $T$. At low $T$ and $P$, a N\'eel antiferromagnetic
insulator~(AFI) appears; it becomes a paramagnetic insulator~(PI) if $T$
increases, or a paramagnetic metal~(PM) if $P$ increases. Dashed line:
first-order transition ending at a liquid-gas critical point. Full lines:
continuous phase transition to the ordered state. Colored regions:
critical~(dark) and mean-field~(light) regimes of the critical point. (b) An
Ising configuration on the triangular lattice. Up~(down) spins correspond to
conducting~(insulating) grains of linear size of the order of the system's
dephasing length.}
\end{figure}

Although band theory predicts that a system of electrons in a solid with one
electron per site (unit cell) should be metallic, such a system ultimately
insulates~\cite{mott49,brinkman70} once the local electron repulsive
interactions exceeds a critical value. The onset of the insulating state,
the Mott transition, arises from the relative energy cost of the on-site
Coulomb repulsion $U$ between two electrons on the same lattice site, and
the kinetic energy, represented by the band width $W$.  Then, the transition
is governed solely by the ratio of $U/W$.  At $T=0$, it is often the case
that symmetries of the microscopic system, associated with charge, orbital
or spin order, may be broken in the Mott insulating state. However, at
sufficiently high temperatures $T$, or in strongly frustrated systems, no
symmetry is broken at the finite-$T$ Mott transition. Then, the transition
is characterized by paramagnetic insulating and metallic phases, whose
coexistence terminates at a second-order critical point, depicted in 
Fig.~\ref{sketchypd}. In this paper, we are concerned with the universal
properties of this \emph{classical critical point}~\cite{comment}, as
revealed by a series of apparently conflicting experiments on 
$(\mathrm{Cr}_{1-x}\mathrm{V}_{x})_{2}\mathrm{O}_{3}$~\cite{limelette03} and organic
salts of the $\kappa -\mathrm{ET}$ family~\cite{kagawa05}. 

Since no symmetry is broken at the finite-$T$ Mott transition, in a strict
sense there is no order parameter. Nonetheless, experimental~\cite{kagawa05,limelette03}, as well as theoretical evidence~\cite{castellani79,kotliar00} suggest that the transition is in the Ising
universality class, similar to the liquid-vapor transition. For example,
Castellani \emph{et al.}~\cite{castellani79} constructed an effective
Hamiltonian for this problem, 
and proposed that double occupancy should play the role of an order
parameter for the Mott transition. On the insulating side, doubly occupied
sites are effectively localized, but in the metal, they proliferate. A
Landau-Ginzburg analysis~\cite{kotliar00} provided further evidence for a
non-analyticity in the double occupancy at a critical value of $U/W$ that
defines a Mott transition. Ising universality follows immediately 
because double occupancy, $\langle n_{i\uparrow }n_{i\downarrow }\rangle $,
is a scalar local density field.

Experimentally, the universality of the Mott critical point is typically
probed by some external parameter, such as pressure, which can tune the
ratio $W/U$. Measurements of the conductivity, $\Sigma $, on 
$(\mathrm{Cr}_{1-x}\mathrm{V}_{x})_{2}\mathrm{O}_{3}$~\cite{limelette03} found that away
from the critical point, the exponents defined through 
\begin{eqnarray}
\Delta \Sigma \left( t,h=0\right) &=&\Sigma (t,h=0)-\Sigma _{c}\propto
|t|^{\beta _{\sigma }} ,  \notag \\
\Delta \Sigma \left( t=0,h\right) &\propto &|h|^{1/\delta _{\sigma }} , 
\notag \\
\partial \Sigma (t,h)/\partial h|_{h=0} &\propto &|t|^{-\gamma _{\sigma }} ,
\end{eqnarray}
have mean-field Ising values, $\beta _{\sigma }\simeq 1/2$,
$\gamma _{\sigma}\simeq 1$ and $\delta _{\sigma }\simeq 3$. 
Here, $t=(T-T_{c})/T_{c} $ and 
$h=(P-P_{c})/P_{c}$, with $(\Sigma _{c},T_{c},P_{c})$ denoting the
corresponding values at the critical endpoint. Close to the critical region,
Limelette \textit{et al.}~\cite{limelette03} observed a drift to the
critical exponents of the 3D Ising universality class. Mean field behavior
is also seen in NiS$_{2}$~\cite{takeshita07}.

However, similar pressure measurements~\cite{kagawa05} on the quasi-2D
organic salts of the $\kappa $-ET family appear to challenge the view that the Mott
transition is in the Ising universality class. In this material, Kagawa 
\textit{et al.}~\cite{kagawa05} found that their data is described by the
exponents $\beta _{\sigma }\simeq 1$, $\gamma _{\sigma }\simeq 1$, and 
$\delta _{\sigma }\simeq 2$, which do not seem to be consistent with the
known exponents of the 2D Ising model whose exponents are~\cite{barrybook} 
$\beta =\frac{1}{8}$, $\gamma =\frac{7}{4}$ and $\delta =15$. Since the
exponents obey the scaling law $\gamma _{\sigma }=\beta _{\sigma }\left(
\delta _{\sigma }-1\right) $, it was proposed that the Mott transition is in
a new, as yet unknown universality class. The situation is further
complicated by thermal expansion measurements~\cite{souza06} that claim to
measure the heat capacity exponent $\alpha $ and find $0.8<\alpha <0.95$.
This result is not only in sharp contrast to the expectation for an Ising
transition (where $\alpha =0$ for $d=2$), it also strongly violates the
scaling law $\alpha +2\beta _{\sigma }+\gamma _{\sigma }=2$, if one uses the
exponents of Ref.~\cite{kagawa05}.

In this paper, we present a unified phenomenological description of all of
these experimental facts within an Ising-type model, and resolve the issue
of the universality class of the Mott transition. A complete description of
these experiments requires to take into account that the conductivity
depends on all possible singular observables of the associated critical
system, and not just on the thermodynamic order parameter associated with
the phase transition. Similar considerations were made in magnetic systems
near the Curie temperature~\cite{hohenberg77, fisher68, mannari68}, to
explain the critical exponent of the conductivity along the coexistence
curve. In that case, a symmetry of the microscopic definition of the
conductance prevented any coupling of the global conductivity to odd moments
of the order parameter, along the coexistence line. Even though similar in
spirit, the situation here is much different. Starting from an effective
microscopic model near an Ising critical point, we show that: 1) the
conductivity typically depends on all possible singular thermodynamic
observables of the system, namely the order parameter and energy density of
the Ising model; 2) when the coupling to the energy density dominates, there
exists a large regime around the critical point, where the critical
exponents for the conductivity are $(\beta _{\sigma },\gamma _{\sigma
},\delta _{\sigma })=(1,\frac{7}{8},\frac{15}{8})$, that agree (within the
error bars) with the findings of Kagawa~\emph{et al.}~\cite{kagawa05}, and the
corresponding mean-field exponents are $(\beta _{\sigma }^{MF},\gamma
_{\sigma }^{MF},\delta _{\sigma }^{MF})=(1,\frac{1}{2},\frac{3}{2})$; 3) a
crossover to Ising exponents is obtained in the order parameter dominated
regime as seen in Refs.~\cite{limelette03,takeshita07}; 4) in the presence
of disorder the Mott critical point ultimately belongs to the random-field
Ising model universality class, and therefore the dimensionality of the
system under study is even more important for specifying its critical
properties.

In order to resolve the discrepancies raised by these experiments, we
consider the behavior of the conductivity of the system near the Mott
critical point, assuming that it belongs to the 2D Ising universality class.
Rather than starting from a microscopic picture, \textit{e.g.\/} a Hubbard
model, we consider a coarse-grained model with the correct symmetries in
which the physics of the relevant transport degrees of freedom is captured.
In this picture, one defines coarse-grained regions, of linear size of the
order of the dephasing length $l_{\phi }$ of the system, which are either
insulating or conducting. Along these lines, we consider an Ising model on a
2D lattice (cf.~Fig.~\ref{lattice}). Near the critical point, where the
correlation length for density fluctuations $\xi $ diverges, it is expected
that the relevant degrees of freedom behave classically. The Ising variables 
$S_{i}$ on each lattice site represent the fluctuating density of mobile
carriers on microscopic ``grains'' of linear size of the order of the
dephasing length $\mathit{l}_{\phi }$, which are conducting ($S_{i}=+1$), or
insulating ($S_{i}=-1 $). The Hamiltonian is 
\begin{equation}
\beta H=-\frac{1}{T}\sum_{\langle ij\rangle }S_{i}S_{j}+\frac{h}{T}
\sum_{i}S_{i} \;,
\end{equation}
where $T$ is the temperature, $P$ and $P_c$ are the pressure and the
critical pressure, respectively, and $h\propto P-P_{c}$ plays the role of
the Ising magnetic field. This model is expected to describe the physics
near the critical point, where $\xi \gg l_\phi$. In this limit, all other
interactions beyond nearest-neighbor are irrelevant. Near the critical
point, the most singular effect of the pressure is described by a coupling
to the order parameter.

To relate the order parameter fluctuations to the transport properties we
will define an associated resistor network for this model, an approach that
has been successfully used in other strongly correlated systems~\cite{carlson-2006,burgy-2001}. Let $\sigma_C$ and $\sigma_I$ be the local
conductivities of the conducting and insulating regions, respectively. We
define the bond conductance of the network model simply by adding these two
conductivities in series. The bond conductance has three possible values,
depending on the state of each grain, which can both be conducting, both
insulating, one conducting and the other insulating. Thus, the conductance
of the bond $(i,j)$ has the form 
\begin{eqnarray}
\sigma_{ij}=\sigma_0\left(1+g_m(S_i+S_j)+g_\epsilon S_iS_j\right)\; .
\label{couplings}
\end{eqnarray}
Even in this toy model, the microscopic conductivity, $\sigma_{ij}$, couples
both to the order parameter, $S_i$, and to the energy density, $S_iS_j$, of
the Ising model with naturally large couplings, $g_m$ and $g_\epsilon$,
defined in Eq.\eqref{couplings}. More specifically, we find that 
$\sigma_0=\frac{1}{4} (\sigma_C+\sigma_I)+\frac{\sigma_C\sigma_I}{\sigma_C+\sigma_I}$, 
$g_m=\frac{ \sigma_C-\sigma_I}{4\sigma_0}$, and $g_\epsilon=\frac{
(\sigma_C-\sigma_I)^2}{4\sigma_0(\sigma_C+\sigma_I)}$. At high contrast, 
$\sigma_C\gg\sigma_I$, we get $g_m\simeq g_\epsilon\simeq1$, whereas, at low
contrast, $|\sigma_I - \sigma_C| \ll \sigma_C$, we get $g_\epsilon < g_m \to
0$.

The conductivity of the 2D Ising model we described is a non-trivial
quantity to compute. As it was shown in the simpler case of the random
resistor network (RRN)~\cite{staufferaharonybook}, networks of bonds with
conductance $\sigma_C$ $(\sigma_I)$ chosen \emph{randomly} with probability 
$p$ and $1 - p$, the global conductivity becomes non-zero as soon as an
infinite percolating and conducting cluster emerges in the system. When 
$\sigma_I=0$, the critical exponent $\beta_\sigma$ of the conductivity is
non-trivially related to the fractal properties of the incipient infinite
conducting cluster. This exponent is larger than unity for random
uncorrelated networks and larger than the exponent of the order parameter,
because dangling bonds of the infinite cluster do not contribute to the
conductivity. On the other hand, it becomes much less than unity for
correlated networks, and typically very close to the exponent of the order
parameter, since the infinite cluster is efficiently connected with few
dangling bonds.

\begin{figure}[tb]
\centering
\includegraphics[width=0.385\textwidth]{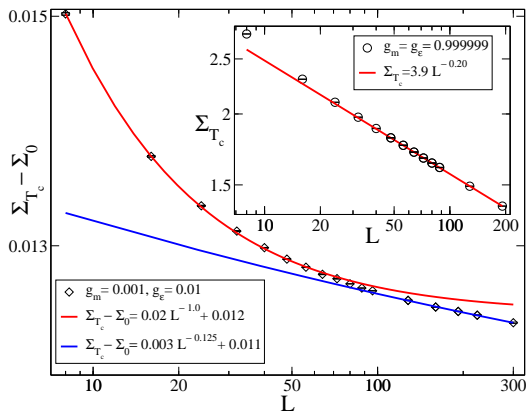}
\caption{Crossover behavior of the conductivity at finite contrast: the
energy density (order parameter) dominates at short (long) length scales.
Inset: Fractal scaling at large contrast.}
\label{expdf}
\end{figure}

On the other hand, when $\sigma_I>0$, a conducting cluster is less
distinguishable from that of an insulating one, and the complex effects
coming from the fractal cluster boundaries are smeared out. In the context
of RRN, the percolation transition is not seen in the behavior of the
conductivity, which seems to show just a crossover. If the contrast is low, 
$\sigma_I\simeq\sigma_C$, the actual conductivity of a single bond between
sites $i,j$, $\Sigma$, should depend only on local observables, and we can
formally expand it in powers of $g_m$ and $g_\epsilon$~\cite{blackman76}, 
\begin{equation}
\Sigma =\sigma_0+g_m\langle (S_i+S_j)\rangle +g_\epsilon\langle S_iS_j
\rangle+\ldots\; ,  \label{sigmaexp}
\end{equation}
where the ellipsis represents more complex products of local spin operators
(weighed by rapidly decaying functions)~\cite{blackman76}. Near the Ising
critical point,  the most singular contribution of the expectation values of
multi-spin operators in Eq.~\ref{sigmaexp} is given by the expectation value
of the most singular, ``primary'', operators of the Ising critical point,
the order parameter $m$ and the energy density $\epsilon$. Thus, the most
singular term of multi-spin operators with odd (even) number of spins is
proportional to the order parameter (energy density). Therefore, within the
range of convergence of this expansion, 
\begin{equation}
\Sigma =\Sigma_{0}(g_{m},g_{\epsilon })+f_{m}(g_{m},g_{\epsilon })\langle
m\rangle +f_{\epsilon }(g_{m},g_{\epsilon })\langle \epsilon \rangle\; ,
\label{Sigmagen}
\end{equation}
where $\Sigma_{0},f_{m},f_{\epsilon }$ are non-universal regular polynomials
in $g_{m}$ and $g_{\epsilon }$. Provided that the critical behavior is still
controlled by the fixed point theory of the Ising model, the total
conductivity should have the structure of Eq.~\eqref{Sigmagen}. Thus, at
finite contrast, Eq.~\eqref{Sigmagen} predicts that the actual conductivity
is the sum of even and odd components, under the Ising symmetry transformation,
$\Sigma=\Sigma_0+\Sigma^{\mathrm{even}}+\Sigma^{\mathrm{odd}}$,
and it should exhibit a \emph{crossover} from an {\em energy density dominated}
behavior at short distances to an {\em order parameter dominated} behavior at
long distances. The crossover scale is controlled by the relative size
of the functions $f_m$ and $f_\epsilon$~(cf.~Fig.~\ref{expdf}). This
behavior breaks down at high contrast where there is multi-fractal behavior
(cf. Inset in Fig.\ref{expdf} and Ref.~\cite{bastiaansen97}).

%%%%%%%%%%%%%%%%%%%%%%%%%%%%%%%%%%%%%%%%%%%%%%%%%%%%%%%%%%%%

\begin{figure}[tb]
\centering
\includegraphics[width=0.4\textwidth]{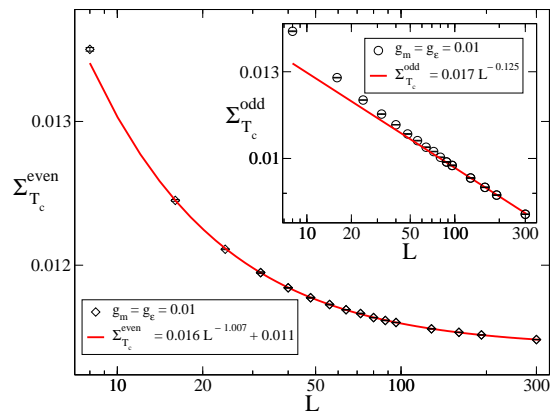}
\caption{Monte Carlo data which verify the expected behavior of the
conductivity when $g_m, g_\protect\epsilon\ll1$. $\Sigma^{\mathrm{even}}$ scales as the energy density (see text). Inset:
$\Sigma^{\mathrm{odd}}$ scales as the order parameter.}
\label{magn_en}
\end{figure}
%%%%%%%%%%%%%%%%%%%%%%%%%%%%%%%%%

We can understand the experiments of Refs.~\cite{kagawa05,limelette03,takeshita07},  
if we assume that Eq.~\eqref{Sigmagen}
applies. The results of Refs.~\cite{limelette03,takeshita07} follow by
assuming that $f_m(g_{m},g_{\epsilon })>f_\epsilon(g_{m},g_{\epsilon })$,
and the conductivity scales as the order parameter. Conversely, the results
of Ref.~\cite{kagawa05} follow if $f_{\epsilon}(g_{m},g_{\epsilon })\gg
f_{m}(g_{m},g_{\epsilon })$,  and the conductivity, for an extended regime
near the critical point, scales as the energy density of the Ising model. In
this case holds $\Delta \Sigma \propto \left\vert m\right\vert ^{\theta }$,
where $\theta =\left( 1-\alpha \right) /\beta $. Then, it follows that 
$\beta _{\sigma }=\theta \beta $, $\delta _{\sigma }=\delta /\theta $ and 
$\gamma _{\sigma }=\gamma +\beta \left( 1-\theta \right) $. The resulting
critical exponents are $(\beta _{\sigma },\gamma _{\sigma },\delta _{\sigma
})=(1,\frac{7}{8},\frac{15}{8})$, very close to the experimental values.
These exponents obey 
\begin{equation}
\gamma _{\sigma }=\beta _{\sigma }\left( \delta _{\sigma }-1\right) \; ,
\end{equation}
if $\gamma =\beta \left( \delta -1\right) $, \textit{i.e.\/}, the
conductivity exponents obey a scaling relation identical to the Ising
exponents, in agreement with the experimental verification of this scaling
relation in Ref.~\cite{kagawa05}. In addition, the scaling function obtained
by Kagawa \emph{et al.}~\cite{kagawa05} only depends on $\beta \delta =\beta
_{\sigma }\delta _{\sigma }$, as in our theory.

In order to verify the theoretical picture presented above, we performed
Monte Carlo simulations of the 2D Ising model on square and triangular
lattices, using the Wolff cluster algorithm~\cite{wolff89}. For the
calculation of the conductivity, for each Ising configuration we used the
Franck - Lobb algorithm~\cite{franck88}, or explicitly solved Kirchhoff
equations. As expected, we found that at the Ising critical point, for 
$g_{m},g_{\epsilon }\ll 1$, the even component of the conductivity 
$\Sigma ^{\mathrm{even}}$ scales as the energy density, while the odd component 
$\Sigma ^{\mathrm{odd}}$ scales as the order parameter~(cf.~Fig.~\ref{magn_en}). As $g_{m},g_{\epsilon }$ approach unity, a slow crossover exists to a
fractal regime of the Ising clusters, which  is crucial for
specifying the critical exponent of the conductivity, consistent with the
results of Ref.~\cite{bastiaansen97} (cf. Inset in Fig.\ref{expdf}.)

Refs.~\cite{limelette03,takeshita07} report 3D mean-field Ising behavior and
a small critical region in 
$(\mathrm{Cr}_{1-x}\mathrm{V}_{x})_{2}\mathrm{O}_{3}$ and $\mathrm{NiS_{2}}$ under pressure respectively, in contrast to the
extended critical region with 2D Ising exponents of Ref.~\cite{kagawa05}. We
can understand these experiments by considering the effects of quenched
disorder on an Ising critical point. The difference between a quasi-2D and a
3D material is a strongly anisotropic Ising interaction along the direction
perpendicular to the planes. Disorder that locally favors the localized over
the delocalized state or vice versa, corresponds to a \emph{random magnetic
field}, and couples to the order parameter of the Ising transition. This
induces density fluctuations. The relevant model for this discussion is the
anisotropic 3D random-field Ising model (RFIM), 
\begin{equation}
H=-J_{xy}\sum_{\{ij\}_{xy}}S_{i}S_{j}-J_{z}\sum_{\{kl\}_{z}}S_{k}S_{l} +
\sum_{i}h_{i}S_{i}\; ,
\end{equation}
where $h_{i}$ is a random field with variance $\Delta $. For $d=3$, there is
a continuous phase transition in the $3D$ random-field Ising model (3DRFIM)
universality class~\cite{nattermann-1998} for any anisotropy $J_{xy}/J_{z}$,
an irrelevant operator at the 3D RFIM fixed point. However, for large
anisotropy and weak disorder, relevant to the quasi-2D organics, there is a
large dimensional crossover regime from 2D RFIM behavior, with an
exponentially long correlation length, to the narrower 3D RFIM criticality~\cite{zachar03}. What changes between the 3D isotropic materials and the
quasi-2D organics is not the universality class, but where the planar
correlations become critical. For weak disorder $\Delta \ll J_{xy} $ and
strong anisotropy $J_{z}/J_{xy}\ll 1$, the planes are essentially decoupled
and 2D-RFIM behavior holds with $\xi _{xy}\gg 1$ in a large region away from
the transition point. When $J_{z}\simeq J_{xy}$, the critical region is
narrow, and controlled by the 3D RFIM fixed point.

With regards to the thermal expansion measurements that claim to measure the
heat capacity exponent $\alpha$, we argue that the authors of 
Ref.~\cite{souza06} misinterpret their results. The volume change is proportional to
the Ising order parameter of the Mott transition, \textit{i.e.\/} $\Delta
l\propto m$, yielding $l^{-1}dl/dT\propto t^{\beta -1}$. The thermal
expansion diverges with exponent $1 - \beta = \allowbreak 0.875$, consistent
with Ref.~\cite{souza06} who find it in the range $0.8-0.95$.

Some major predictions can be drawn from our picture. Firstly, all
thermodynamic observables near the Mott critical point should have Ising
critical exponents. Secondly, regarding the critical behavior in quasi-2D
organic salts~\cite{kagawa05}, in the clean system, the conductivity along
the coexistence line should have the same critical exponent ($\beta _{\sigma
}=1$) in both mean-field and true-critical regimes. This means that the
conductivity jump $\Delta \Sigma _{J}\equiv \Sigma (T,h=0^{+})-\Sigma
(T,h=0^{-})$ along the coexistence line, which should be proportional to the
order parameter, should have distinct mean-field and critical regimes, where 
$\beta _{\Delta \Sigma _{J}}=1/2$ and $\beta _{\Delta \Sigma _{J}}=1/8$
respectively. Also, the first-order Mott transition is expected to be
broadened by disorder~\cite{imry-wortis-1979,aizenman-wehr-1989}. Thus,
instead of a sharp jump in the conductance one should see a continuous
change which would become more abrupt for clean systems. The net effect is
to make the system spatially inhomogeneous, as in charge ordered phases,
stripes and electron nematics~\cite{kivelson-1998} and in the manganites~
\cite{dagotto-2001}, which tend to round their phase transitions and replace
the first-order transition by an inhomogeneous phase. Thus, hysteretic
glassy-like aging effects~\cite{sethna-2004}, a problem that has been
studied only recently in strongly correlated systems~\cite{schmalian-wolynes-2000,carlson-2006}, are also expected at this Mott
transition.

In conclusion, we have explained under a consistent phenomenological
framework the series of experiments that were performed during the last few
years on Mott criticality. We showed that the conductivity of a system near
a critical point depends on all possible local singular observables of the
system, which in the case of interest, are the order parameter and the
energy density of the effective Ising model. This description holds when the
contrast between conducting and insulating regions is small. Should this not
hold, fractal behavior of the incipient infinite clusters at the critical
point is crucial for specifying the critical exponents of the conductivity. 
Finally, disorder affects the effective dimensionality of the system. In
particular, we showed that for quasi-2D materials, such as the organic salts
in the $\kappa $-ET family, critical fluctuations are expected to be much
larger than in a 3D material in an extended regime in the $(P,T)$ plane
around the Mott critical point.

\paragraph{Acknowledgements}

We thank K. Kanoda for discussions. This work was supported in part by the
National Science Foundation grants DMR 0442537 (EF) and DMR 0605769 (PP) at
UIUC,  by the U.S. Department of Energy, Division of Materials Sciences
under Award DE-FG02-07ER46453 (EF), through the Frederick Seitz Materials
Research Laboratory at UIUC, and
the Ames Laboratory,  operated for the U.S. Department of Energy by Iowa
State University under Contract No. DE-AC02-07CH11358 (JS), and by
CAPES and CNPq (Brazil) (RF).

\end{document}